\pdfoutput=1

\documentclass[%
reprint,
superscriptaddress,
longbibliography,
aip,
showpacs]{revtex4-1}

\usepackage{graphicx}
\usepackage{amsmath,amsfonts,
}

\usepackage{epstopdf}
\usepackage{subfigure}

\usepackage{amssymb}
\usepackage{bm}
\usepackage{lineno}

\begin{document}

\title{Macroscopic acousto-mechanical analogy of a microbubble}

\author{Jennifer~Chaline}
\affiliation{UMR Inserm U930, Universit\'{e} Fran\c{c}ois-Rabelais, Tours, France}

\author{No\'e~Jim\'enez}
\affiliation{Instituto de Investigaci\'on para la Gesti\'on Integrada de Zonas Costeras, Universitat Polit\`ecnica de Val\`encia, Paranimf 1, 46730 Grao de Gandia, Spain}

\author{Ahmed~Mehrem}
\affiliation{Instituto de Investigaci\'on para la Gesti\'on Integrada de Zonas Costeras, Universitat Polit\`ecnica de Val\`encia, Paranimf 1, 46730 Grao de Gandia, Spain}

\author{Ayache~Bouakaz}
\affiliation{UMR Inserm U930, Universit\'{e} Fran\c{c}ois-Rabelais, 10 boulevard Tonnellé, 37032 Tours, France}

\author{Serge~Dos~Santos}
\affiliation{INSA Centre Val de Loire, 3 rue de la Chocolaterie CS 23410, F-41034 Blois, France}
\affiliation{UMR Inserm U930, Universit\'{e} Fran\c{c}ois-Rabelais, Blois, France}

\author{V\'{i}ctor J.~S\'anchez-Morcillo}
\affiliation{Instituto de Investigaci\'on para la Gesti\'on Integrada de Zonas Costeras, Universitat Polit\`ecnica de Val\`encia, Paranimf 1, 46730 Grao de Gandia, Spain}

\date{\today}

\begin{abstract} 
Microbubbles, either in the form of free gas bubbles surrounded by a fluid or encapsulated bubbles used currently as contrast agents for medical echography, exhibit complex dynamics under specific acoustic excitations. Nonetheless, considering their micron size and the complexity of their interaction phenomenon with ultrasound waves, expensive and complex experiments and/or simulations are required for their analysis. The behavior of a microbubble along its equator can be linked to a system of coupled oscillators. In this study, the oscillatory behavior of a microbubble has been investigated through an acousto-mechanical analogy based on a ring-shaped chain of coupled pendula. Observation of parametric vibration modes of the pendula ring excited at frequencies between $1$ and $5$ Hz is presented. Simulations have been carried out and show mode mixing phenomena. The relevance of the analogy between a microbubble and the macroscopic acousto-mechanical setup is discussed and suggested as an alternative way to investigate the complexity of microbubble dynamics.

\end{abstract}

\pacs{43.25.Ts,43.20.Wd,43.25.Gf}

\maketitle

\section{Introduction}

When subjected to an external acoustic field, bubbles can undergo complex radial oscillations. This oscillatory behavior has been an important and continuously developing research subject since the beginning of the twentieth century. The investigation of bubble dynamics started with the work of Lord Rayleigh \cite{Rayleigh1917}, who was mandated by the Royal Navy to explain the origin of damages on submarine propellers. Rayleigh focused on the oscillatory behavior of cavitation bubbles suspended in a fluid. He showed that the overpressure generated by the oscillations and the collapse of bubbles could explain the damages caused on propellers. In the 30's, Minnaert was interested in the origin of the sound of running water \cite{Minnaert1933}. He supposed that bubbles oscillating periodically in water were at the origin of rivers whispering. After these seminal works, many studies on bubble oscillations have been carried out in different research fields. 

In addition to the radial motion, bubbles can show non-spherical oscillations or vibration modes. These vibration modes, characterized by an index $n$, were first analyzed theoretically at the interface between immiscible and incompressible fluids with spherical symmetries \cite{Plesset1954}. Later, Neppiras \cite{Neppiras1969} analyzed the acoustic response from gas bubbles suspended in a fluid and subjected to sound fields. Eller and Crum \cite{Eller1970}, and Prosperetti \textit{et al.} \cite{Prosperetti1988} focused on the instability of the motion and the nonlinear dynamics of a bubble within a sound field. In the 90's, the discovery of single bubble sonoluminescence (SBSL) by Gaitan and Crum \cite{Gaitan1992}, led to additional studies on  nonlinear oscillations of bubbles \cite{Brenner1995,Gaitan1998,Augsdoerfer2000}. Finally, with the use of ultrasound contrast agents \cite{Ophir1989}, the understanding of bubble dynamics have found a renewed interest in the field of ultrasound imaging and targeted drug delivery \cite{Escoffre2011,Novell2013}. Therefore, the problem of vibration modes in bubbles is still under investigation.

In this study, the interaction between ultrasound and a microbubble, and especially the appearance of vibration modes, are studied using a macroscopic analog system. The mechanisms underlying its nonlinear behavior are sought to improve our understanding of the microbubble dynamics. The study of a single microbubble is a difficult task, particularly because of the interaction with other microbubbles or microstreaming that sweep away the bubble. Moreover the smallness and the complexity of the phenomena involved, require complex modeling and expensive experiments. In this work, we propose the use of a macroscopic mechanical analog as an alternative way to investigate microbubble dynamics.

The study of analog models in the field of physics is a tool that allows recreating in the laboratory phenomena that are difficult to observe directly. Concerning microbubbles, some of its dynamic features (vibration modes, subharmonic oscillations, chaos) can be captured by a system of coupled oscillators. Here, the oscillatory behavior of a microbubble and its vibration modes are investigated through the use of a macroscopic analogy consisting of a chain of coupled pendula, parametrically excited by a vertical force. Based on a discrete nonlinear model of coupled pendula and its continuous limit describing low frequency excitations, vibration modes are investigated theoretically and experimentally. This approach is used here to set up a formal basis for the acousto-mechanical analogy of a gas microbubble.

The structure of the paper is as follows: Mathematical formulations of a gas microbubble in an ultrasound field, and the analog acousto-mechanical system are given and compared in sections II and III respectively. The experimental setup and the results of the measurement of vibration modes in the macroscopic system are described in section IV, where numerical simulations are also presented. The conclusions of the study are presented in section V.

\section{Microbubble Dynamics}

\begin{figure}[b]
\begin{center}
\includegraphics[width=8cm]{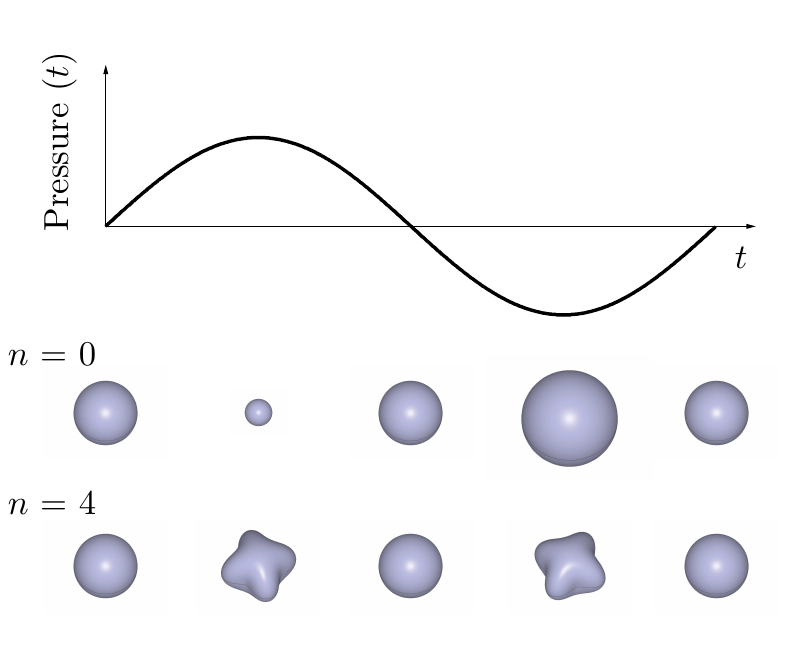}
\caption{A bubble oscillating volumetrically within an acoustic field. For positive and negative pressures, compression and expansion phases are observed respectively.}
\label{fig:linear}
\end{center}
\end{figure}

In the presence of an acoustic field, microbubbles can be forced to oscillate in different ways.
The most common oscillation mode is the radial mode (Fig.~\ref{fig:linear}), where the bubble compresses
and expands radially, maintaining its spherical shape. The basic model describing the radial
dynamics of a bubble is the Rayleigh-Plesset equation for the time dependent radius $R(t)$
\begin{equation}\label{RP}
	\rho \left(R \ddot{R}+\frac{3}{2} \dot{R}^2 \right) =P_g-P_0-P_A(t)-\frac{2 \sigma}{R}-4\mu \frac{\dot{R}}{R},
\end{equation}

\noindent with $\rho$ the density of the surrounding fluid, $P_g=\left(P_0+\frac{2 \sigma}{R_0} \right) \left(  \frac{R_0}{R} \right)^{3\gamma}$, the gas pressure inside the bubble, $P_0$ is the hydrostatic pressure, $R_0$ is the equilibrium radius of the bubble, $P_A$ the acoustic pressure, $\sigma$ the surface tension of the bubble, $\mu$ the dynamic viscosity and $\gamma$ the polytropic exponent. Some generalizations of this model have been proposed \cite{Church1995,Hoff2000,Chatterjee2003,Marmottant2005}. The resonance frequency of the radial mode of the bubble, the so called Minnaert frequency, can be obtained from Eq. (\ref{RP}) and is given by 

\begin{equation}
f = \cfrac{1}{2\pi R_0}\left(\cfrac{3\gamma P_0}{\rho}\right)^{1/2}.
\label{Minnaert}
\end{equation}

 Under typical conditions, the relation $fR_0$ is a magnitude of order 1. In particular, for a bubble in water at standard pressure ($P_0$=100 kPa, $\rho$=1000 kg/m$^3$), this equation gives the condition $f R_0 \approx 3.26$~ m/s. 

\begin{figure}[t]
\begin{center}
\includegraphics[width=8cm]{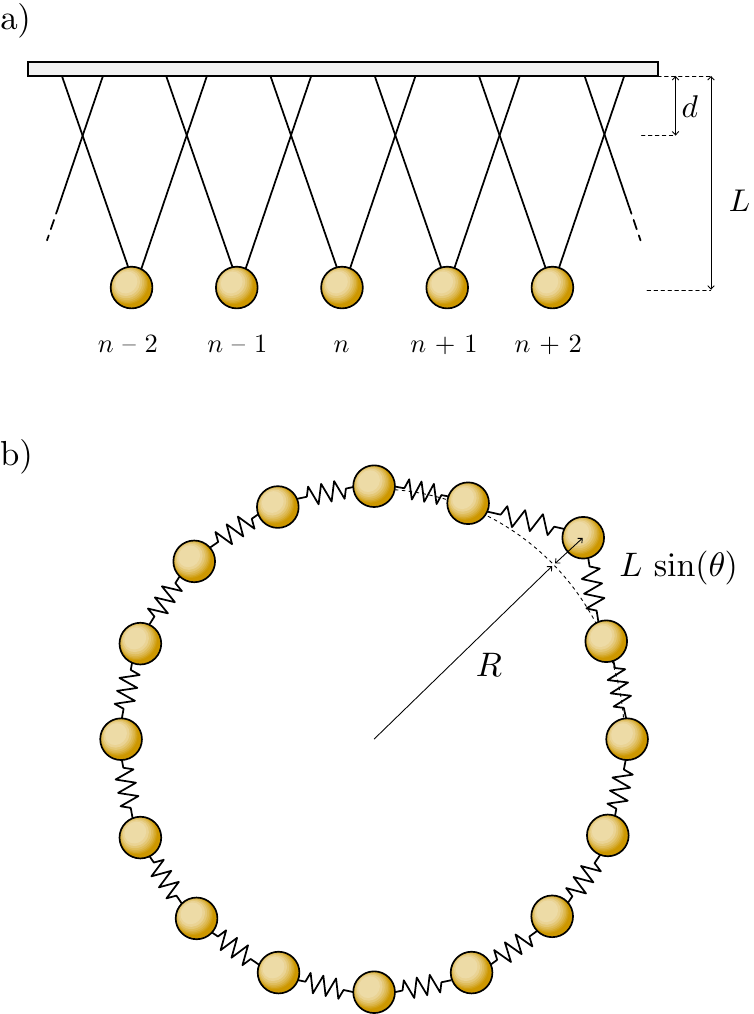}
\caption{Representation of microbubble surface modes with $n$ the mode's order. The left pictures are experimental results from \citet{Versluis2010}, and show a selection of surfaces modes observed for different microbubble radius between 36-45 $\mathrm{\mu}$m. The two columns in the center are 3D analytical solutions from Eq. (\ref{harmonicsdef}), and its equatorial cross-section. The right column shows the corresponding vibration modes experimentally observed with the acousto-mechanical system, as described in Section IV.}
\label{fig:vibration}
\end{center}
\end{figure}

Microbubbles can also undergo non-spherical oscillations (see Fig.~\ref{fig:vibration}, left column) through instabilities at the gas/fluid interface. In this case, the radius becomes a space-dependent function,
\begin{equation}\label{radiusmodified}
	R(t)\rightarrow R(\theta,\varphi,t)=R(t)+\xi(\theta,\varphi,t),
\end{equation}
\noindent where $R(t)$ describes the evolution of the radial mode and $\xi(\theta,\varphi,t)$ a perturbation depending on the spherical coordinates $\theta$ and $\phi$. As Fig.~\ref{fig:vibration}~(left) shows, microbubbles can develop different surface patterns that depend on the excitation parameters (amplitude and frequency of the ultrasound wave) and the bubble radius. The radial oscillation corresponds to the mode $n=0$; the mode $n=1$ corresponds to the displacement of the center of mass. Non-spherical surface modes are those modes with $n\geq 2$.

To describe non-spherical modes in bubbles, a common analytical approach is to expand the perturbation of the radial mode, $\xi(\theta,\varphi,t)$, on the basis of spherical harmonics \cite{Plesset1954}
\begin{equation}\label{harmonics}
	\xi(\theta,\varphi,t)=\sum_{n,m}a_n(t)Y^m_n(\theta,\varphi), 
\end{equation}

\noindent where $a_n(t)$ is the time-dependent amplitude of the surface mode of index $n$, and $Y^m_n(\theta,\varphi)$ is the spherical harmonic defined as
\begin{equation}\label{harmonicsdef}
	Y^m_n(\theta,\varphi)=\frac{(-1)^m}{ \sqrt{4\pi}}\sqrt{\frac{(n-m)!}{(n+m)!}}\sqrt{2n+1}P_n^m(\cos\theta)e^{im\varphi}, 
\end{equation}
\noindent where $P_n^m$ are Legendre polynomials. Although $Y^m_n(\theta,\varphi)$ defines a large set of possible surface modal oscillations, experiments \cite{Versluis2010} show that the observed modes present symmetry along the axis of the incident ultrasound beam, corresponding to $m=0$, also known as zonal harmonics. Then, spherical harmonics $Y_n^0(\theta,\phi)$ reduce to Legendre polynomials $P_n(\cos\theta)$. The evolution equation for the amplitude of each mode can be found by matching velocity potentials and pressures at both sides of the interface, and linearizing for small amplitudes \cite{Plesset1954}, 
\begin{eqnarray}
&&\ddot{a}_n+\left(\frac{3\dot{R}}{R}+\frac{2(n+2)(2n+1)}{\rho R^2}\mu \right)\dot{a}_n+ \nonumber \\
&&\left(\frac{(n+1)(n+2)\sigma}{\rho R^3}+\frac{2(n+2)\mu \dot{R}}{\rho R^3}-\frac{\ddot{R}}{R}\right)(n-1)a_n=0,
\label{amplitude}
\end{eqnarray}
\noindent where, $\mu$ is the viscosity and $\sigma$ the surface tension. Defining $b(t)=a(t)R^{3/2}$, the equation can be simplified as \cite{Leighton1994}
\begin{equation}
\ddot{b}_n+\left(\frac{(n-1)(n+1)(n+2)\sigma}{\rho R^3}-\frac{3\dot{R}^2}{4R^2}-\frac{(2n+1)\ddot{R}}{2R}\right)b_n=0.
\label{simplify}
\end{equation}

Within this approach, each mode $n$ obeys to the equation of a harmonic oscillator, with time dependent coefficients. The resonance frequencies of the surface modes readily follow from Eq.~(\ref{simplify}) by considering the static condition $R=R_0$, and $\dot{R}=\ddot{R}=0$. 
\begin{equation}
\omega_n=\sqrt{\frac{(n-1)(n+1)(n+2)\sigma}{\rho R_0^3}},
\label{frequencies}
\end{equation}

\noindent which is the Lamb expression for surface modes for a free gas bubble.

The acoustic pressure term $P_A$ in Eq.~(\ref{RP}) is usually a harmonic function with angular frequency $\omega_e$, i.e. $P_A(t)=p_A \cos(\omega_e t)$. For sufficiently small amplitudes $p_A$, the bubble response will be also harmonic at the same frequency, i.e.
\begin{equation}
R(t)=R_0+R_\varepsilon\cos(\omega_e t),
\label{rayon}
\end{equation}

\noindent with $R_\varepsilon \ll R_0$. Substituting Eq.(\ref{rayon}) into Eq.(\ref{simplify}) and linearizing, a Mathieu equation for each surface mode is obtained \cite{Leighton1994}

\begin{equation}
\ddot{b}_n+\left(\omega_n^{2}+\left(\frac{(2n+1)\omega_e^2}{2}-3\omega_n^2\right)\frac{R_\varepsilon}{R_0}\cos(\omega_e t)\right)b_n=0.
\label{Mathieu}
\end{equation}

The Mathieu equation is a special case of a linear second-order homogeneous differential equation with time dependent coefficients, and appears in many applications in physics and engineering, specially in the description of parametrically driven systems \cite{Ruby1996}, as the systems considered in this work.

\section{The parametrically driven chain of coupled pendula}

In this section the model equations for the macroscopic analog of the microbubble are formulated. First, we consider the exact problem of the discrete lattice of coupled masses, that corresponds to our experimental system. Later, the continuum limit of this model is used to establish the analogy with gas microbubble, by deriving an equation isomorphic to Eq. (\ref{Mathieu}). 
\begin{figure}[tb]
	\begin{center}
	\includegraphics[width=7cm]{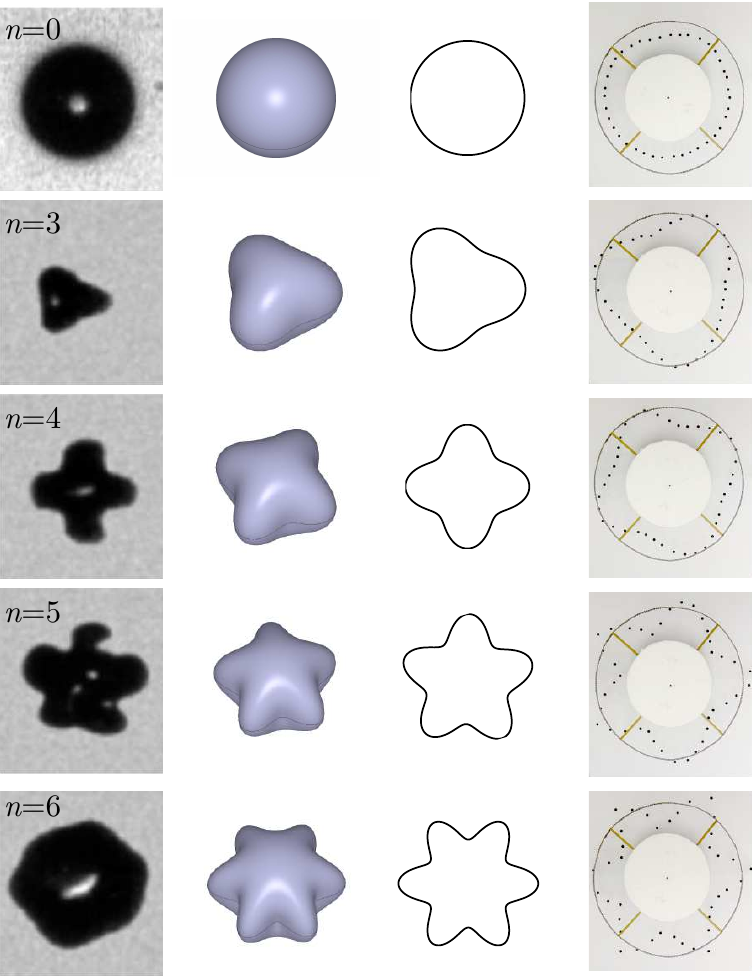}
	\caption{Schematic representation of the chain of pendula coupled by V-shaped strings and knots, hanging from a rigid support. (a) Lateral view of a section of the chain, (b) top view of the whole chain in a circular arrangement. Deviation of one mass with respect to its equilibrium position, and its coordinates are shown for illustration.}
	\label{fig:scheme_lattice}
	\end{center}
\end{figure}

\subsection{The discrete lattice}
The equation of motion of a pendulum of length $L$ is given by 
\begin{equation}
\ddot{\Theta}+\omega_0^2\sin \Theta=0,
\label{pendulum}
\end{equation}
where $\Theta(t)$ is the angle with respect to the vertical, and $\omega_0=\sqrt{g/L}$, $g$ being  the acceleration due to gravity. Here, for the sake of simplicity we have ignored dissipation, which can be included in Eq.~(\ref{pendulum}) by an additional term $\beta \dot{\Theta}$. When a set of oscillators are coupled to their nearest neighbors, they form a lattice or chain, supporting waves. Consider that the lattice is subjected to a parametric forcing with displacement amplitude $h_e$ and angular frequency $\omega_e$, then its motion is described by  
\begin{equation}
\ddot{\Theta}_i+(\omega_0^2+\eta \cos\omega_et)\sin \Theta_i-c^2(\Theta_{i+1}-2\Theta_i+\Theta_{i-1})=0,
\label{motion}
\end{equation}
where $\Theta_i$ is the angle of the $i$-th pendulum, $\eta=4\omega_e^2h_e/L$ is the forcing parameter, and $c$ is a constant denoting the strength of the coupling, equivalent to the speed of sound for the waves for the lattice waves. Lattice models like Eq.~(\ref{motion}) have been extensively studied, in the context of Frenkel-Kontorova chain \cite{Kivshar04}. Most of the work is theoretical, and focused on the formation of localized states (breathers, solitons and kinks). There are few experimental works \cite{Denardo92,Chen94,Thakur08}, where different parametric driving mechanisms and coupling have been implemented. In Ref. \cite{Denardo92}, coupling is achieved by strings and knots, while in Ref. \cite{Thakur08} torsional springs are used. In all the cases, a linear arrangement of pendula was considered. Boundary conditions typical of this configuration are either fixed or free ends.

Here, in order to mimic a bubble-like behavior, we consider the chain in a circular arrangement. This implies the periodic boundary condition $\Theta_{N+1}=\Theta_1$. In addition to this, the circular configuration incorporates curvature effects, which are not present in the linear chain. We assume an V-shaped coupling (see Fig. \ref{fig:scheme_lattice}(a)), for which coupling strength is given by 
\begin{equation}
	c^2=\frac{gda^2}{4L(L-d)},
\label{coupling}
\end{equation}

\noindent where $d$ is the distance from the knot to the support, and $a$ the distance between masses \cite{Denardo92}. Note that, in this model, the coupling strength can be varied by selecting the position of the knot.	

The chain supports different oscillation modes, labeled with an integer index $n$, corresponding to standing waves with wavenumber $k_n$. For the circular chain of radius $R_0$, the relation $k_n=n/R_0$ holds. The relation between driving frequency and mode index (the dispersion relation) can be obtained, for negligible damping and forcing, after linearization and assuming a solution in the form of a harmonic wave, $\exp i(kna-\omega t)$. This results in the relation
\begin{equation}
\omega_n=\sqrt{\frac{g}{L-d}\left[1-\frac{d}{L}\cos^2\left(\frac{na}{2R_0}\right)\right]}.
\label{dispersion}
\end{equation}

As shown in Fig. \ref{fig:scheme_lattice}(b), which illustrates the chain viewed from the top, the ring of oscillators is equivalent to an equatorial section of a bubble with equilibrium radius $R_0$, 
\begin{equation}
R_i(t)=R_0+L\sin\Theta_i(t)
\label{radius}
\end{equation}

\noindent where $R_0$ is the radius of the ring (equivalent to the equilibrium radius of the bubble).
Assuming small angles $\sin\Theta_i\simeq\Theta_i$, the position of each pendulum at a time $t$ can be given by $R_i(t)=R_0+L\Theta_i(t)$
 
\subsection{Continuous description}
 
The bubble surface is a continuum, while the chain of pendula is a discrete system. Therefore, analogies must be searched in the limit where a continuum version of Eq. (\ref{motion}) applies. This is the case when we restrict our analysis to long-wavelength modes (low index $n$), where the mode scale are much larger than the distance between two pendula, $k_n a<<1$, or $na/R_0<<1$. Considering this limit, the discrete angle coordinates $\Theta_i(t)$ can be replaced by the continuous function $\Theta(x,t)$. 

 The equation of motion Eq.~(\ref{motion}) leads to the parametrically driven, small amplitude,
sine--Gordon model, that reads

\begin{equation}
\ddot{\Theta}-c^2 \Theta_{xx}+(\omega_0^2+\eta \cos(\omega_e t))\Theta=0
\label{motion2}
\end{equation}
\noindent where $\Theta$ is a surface deformation related to the bubble radius as $\Theta=(R-R_0)/L$.

Solutions of Eq. (\ref{motion2}) can be expressed as a superposition of normal modes. For a linear (straight) chain, a proper basis is given by the harmonic functions 
 \begin{equation}
 \Theta(x,t)=\sum_n{b_n(t)\cos(k_nx)}.
 \label{solution}
 \end{equation}
\noindent Some solutions of Eq. (\ref{motion2}) have been discussed for a one-dimensional straight geometry \cite{Armbruster01}.

In the ring geometry considered here, angular coordinates are more suitable to describe the position of a point on the deformed ring, and we use the transformation $x=R\theta$, where $\theta=[0,2\pi[$ and $R=R_0$ is the equilibrium radius. The Laplacian operator takes the form $\Theta_{xx}=R^{-2} \Theta_{\theta\theta}$. A proper basis for the expansion in this case is 
 \begin{equation}
 \Theta(\theta,t)=\sum_n{b_n(t) P_n(\cos\theta)},
 \label{solution2}
 \end{equation}
\noindent where $P_n(\cos\theta)$ is a Legendre polynomial and $\theta$ is the angular coordinate.

The equation for the temporal evolution of the mode amplitudes can be obtained by substituting  Eq.~(\ref{solution}) or Eq.~(\ref{solution2}) into Eq.~(\ref{motion2}) and projecting over the different modes (Galerkin projection) using the orthogonality properties of harmonic functions or Legendre polynomials. Independently of the basis chosen for the expansion, the following equation is obtained
\begin{equation}
\ddot{b}_n+\left(\omega_n^2+\eta \cos(\omega_e t)\right)b_n=0,
\label{equapendula}
\end{equation}
\noindent which is a Mathieu equation, where the parametric excitation amplitude is $\eta=\frac{4\omega^2_e h_e}{L}$ and the frequency on the $n$-th mode is given by
\begin{equation}
\omega_n^{2}=\omega_0^2+c^2\frac{n^2}{R_0^2}.
\label{pulsationmode}
\end{equation}
The same result is obtained from the exact dispersion relation Eq.~(\ref{dispersion}) for the discrete system, evaluated in the limit of low index $n$, where $na/R_0<<1$.

\subsection{Analogy}

\begin{table}[b]
\centering\small
	\caption{Table of analogies between parameters of microbubble and those of the pendula ring.}
	{\begin{tabular}{@{}lcc@{}}
\toprule
		{ Parameter}  & { Microbubble} & { Pendula ring}  \\
\hline
		$R_0$ [m] & $\simeq 10^{-6}$  & $\simeq 0.3$  \\
		$f_e$ [Hz]  & $\simeq 10^6$  & $\simeq 3$   \\
		$f_e\cdot R_0$  & 1 & 1\\
		$\omega_n^2$     	 & $(n-1)(n+1)(n+2)\frac{\sigma}{\rho R_0^3}$& $\frac{g}{L}+n^2\frac{c}{R_0^2}$ \\
		$f_p$ & $\frac{(2n+1)}{2}\omega_e^2\left(\frac{R_\varepsilon}{R_0}\right)$ & $4\omega_e^2\left(\frac{A}{L}\right)$ \\
\hline
	\end{tabular}}\label{table_analogy}
\end{table}

The previous theoretical study stablish a number of analogies between the ultrasound driven gas bubble and the ring of coupled pendula subjected to a time-dependent acceleration. Both systems are described by the same Mathieu equation, with different coefficients. A similar behavior is therefore expected for both systems, and similarity parameters are listed in table \ref{table_analogy}. For typical experiments, microbubbles ($R_0\simeq 10^{-6}$ m) are insonified in the MHz frequency range. Note that, from Eq. (\ref{Minnaert}), the product $f R_0$ is fixed in microbubbles, but can be tuned in the mechanical macrobubble, where $f$ and $R_0$ can be varied independently.  The pendula ring used here, with $R_0\simeq 0.3$ m, is designed to be excited in the Hz frequency range. We can therefore consider $f \cdot R_0$ as a geometrical invariant.

\section{Experiments}

	\subsection{Experimental setup}
\begin{figure}[b]
	\begin{center}
	\includegraphics[width=6cm]{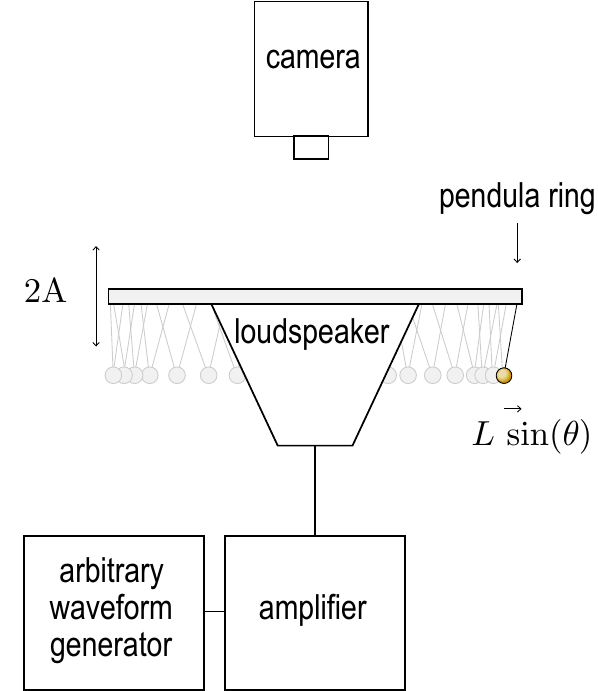}
	\caption{Scheme of the experimental setup. The ring of coupled pendula is driven vertically by an sinusoidal force generated by a loudspeaker.}
	\label{fig:scheme_montage}
	\end{center}
\end{figure}

The setup consists of an aluminum ring, on which pendula of mass $m=6$ g are fixed with nylon strings forming a ``V'' shape with the vertical axis as shown in Fig.~\ref{fig:scheme_lattice}(a). The coupling between pendula is obtained by overlapping the strings and fixing them by a knot \cite{Denardo92}. The parameter with a stronger influence on the dynamics of the chain is the coupling strength. Here, we focus on a ring with a medium coupling. The ring has a radius $R=31$ cm, $N=54$ pendula, pendulum's length is $L=10$ cm and the ldistance between the ring and the node is $d=5$ cm. As Fig.~\ref{fig:scheme_montage} shows, the pendula ring lies on the excitation system.
The mechanical excitation system consists of a subwoofer loudspeaker driven by an arbitrary waveform generator (Agilent33220A) through an audio amplifier. Thus, the pendula ring is attached to the subwoofer cone and therefore is excited mechanically by a vertical oscillatory force, as shown in Fig.~\ref{fig:scheme_montage}. The sinusoidal excitation varies from $f_e=0$ to $5$  Hz with amplitudes varying from $A=0.5$ to $3.5$ V$_\mathrm{pp}$, corresponding to a vertical displacements ranging from $h_e=0.5$ mm to $3.5$ mm. The pendula are driven near the volumetric resonant frequency and its double to allow the development of surface modes during the mechanical excitation through parametric instability.

The motion of pendula has been recorded from the top with a video camera. The data processing has been performed with the software Image J and the plug-in MJ Track. This plug-in enable to track the motion of each pendulum and thus to determine the distance $R_i(t)$ between the center of the ring and a pendulum at a fixed time $t$. Applying a spatial fast Fourier transform on $R_i(t)$, the amplitude of each vibration mode of order $n$ can be obtained.

\subsection{Results}
\begin{figure}[tb]
	\begin{center}
	\includegraphics[width=4cm]{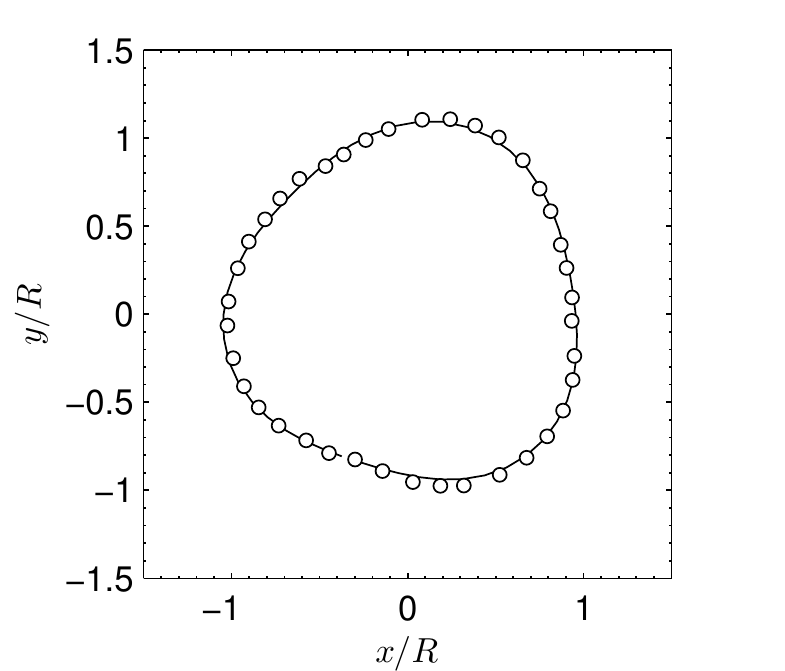} \includegraphics[width=4cm]{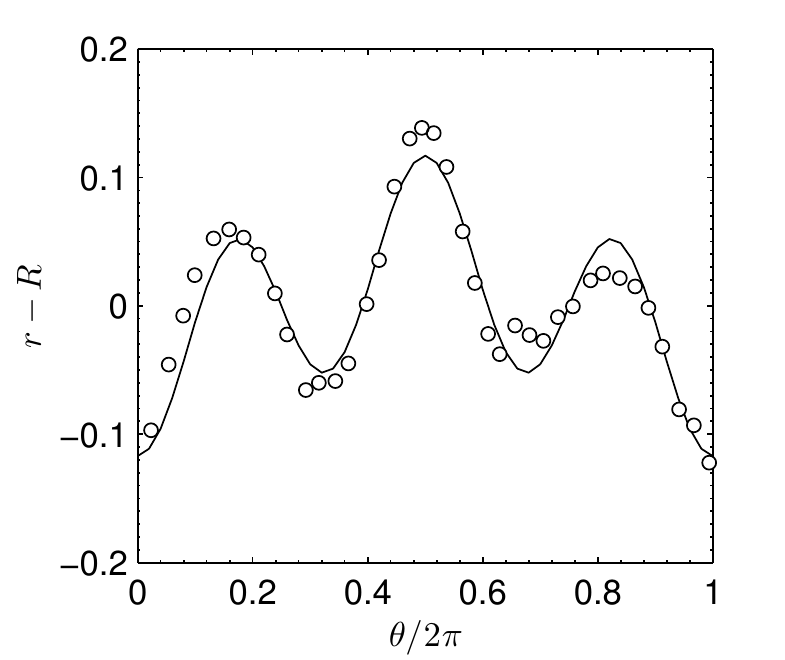}
	\caption{Experimental mode $n=3$ (symbols) and the corresponding fit to a Legendre polynomial $P_3(\cos\theta)$. Two different representations, in cartesian (left) and polar (right) coordinates}
	\label{fig:mode3}
	\end{center}
\end{figure}

A set of measurements has been carried out for excitation amplitudes between $h_e=0.5$ and $3.5$ mm and frequencies varying from $f_e=0$ to $5$ Hz. Vibration modes have been observed up to $n=22$,  including an unstable volumetric mode for $n=0$. The mode $n=1$ corresponding to the displacement of the center of mass has also been observed. An example of the observed patterns is given in Fig.~\ref{fig:mode3} for an excitation amplitude $h_e=2$ mm and frequency $f_e=3.20$ Hz. Here, the pendula ring shows a mode $n=3$. A fit to the corresponding Legendre polynomial, shown in continuous line, shows a good agreement. 

\begin{figure}[tb]
	\begin{center}
	\includegraphics[width=8cm]{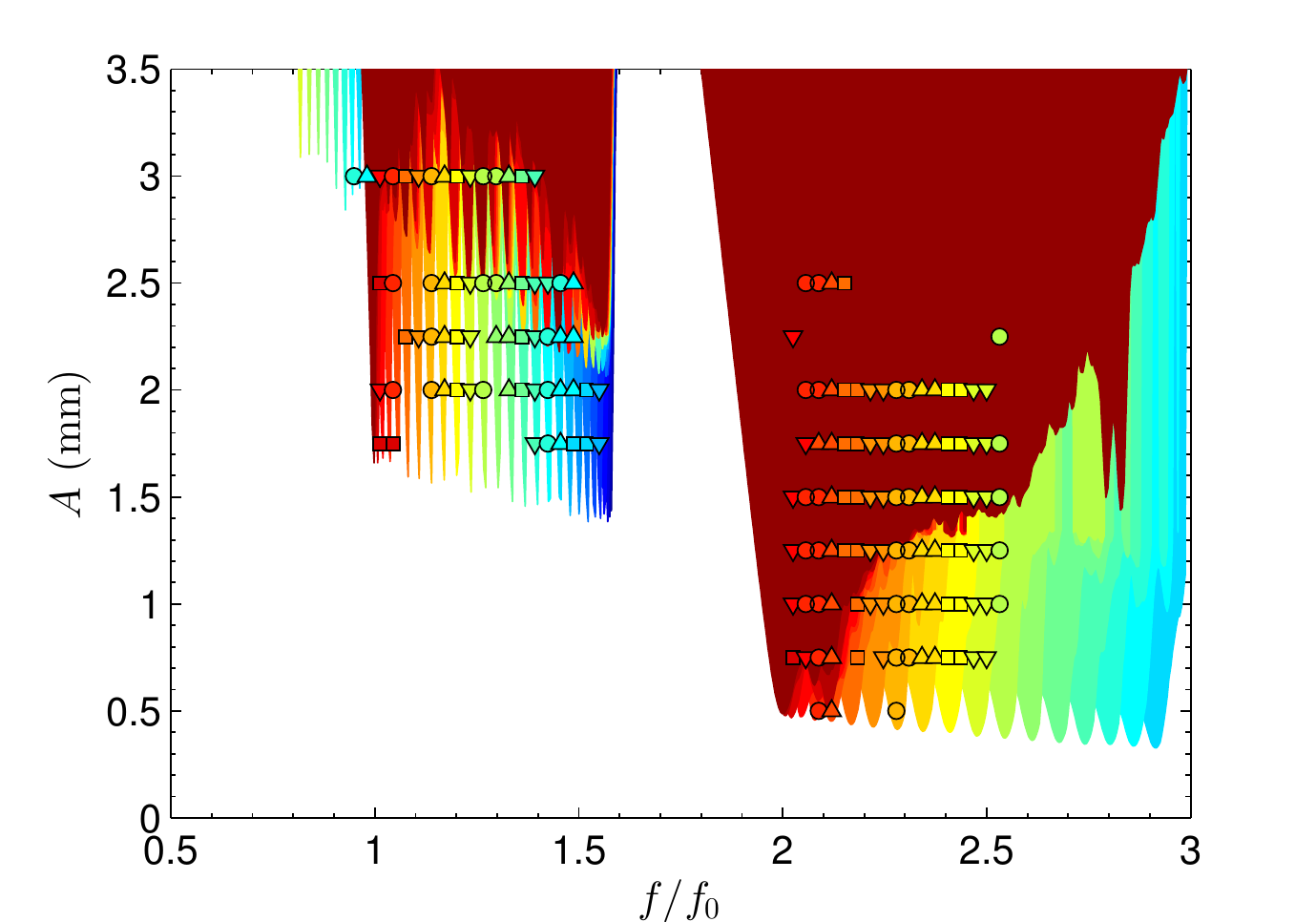} \\
	\includegraphics[width=8cm]{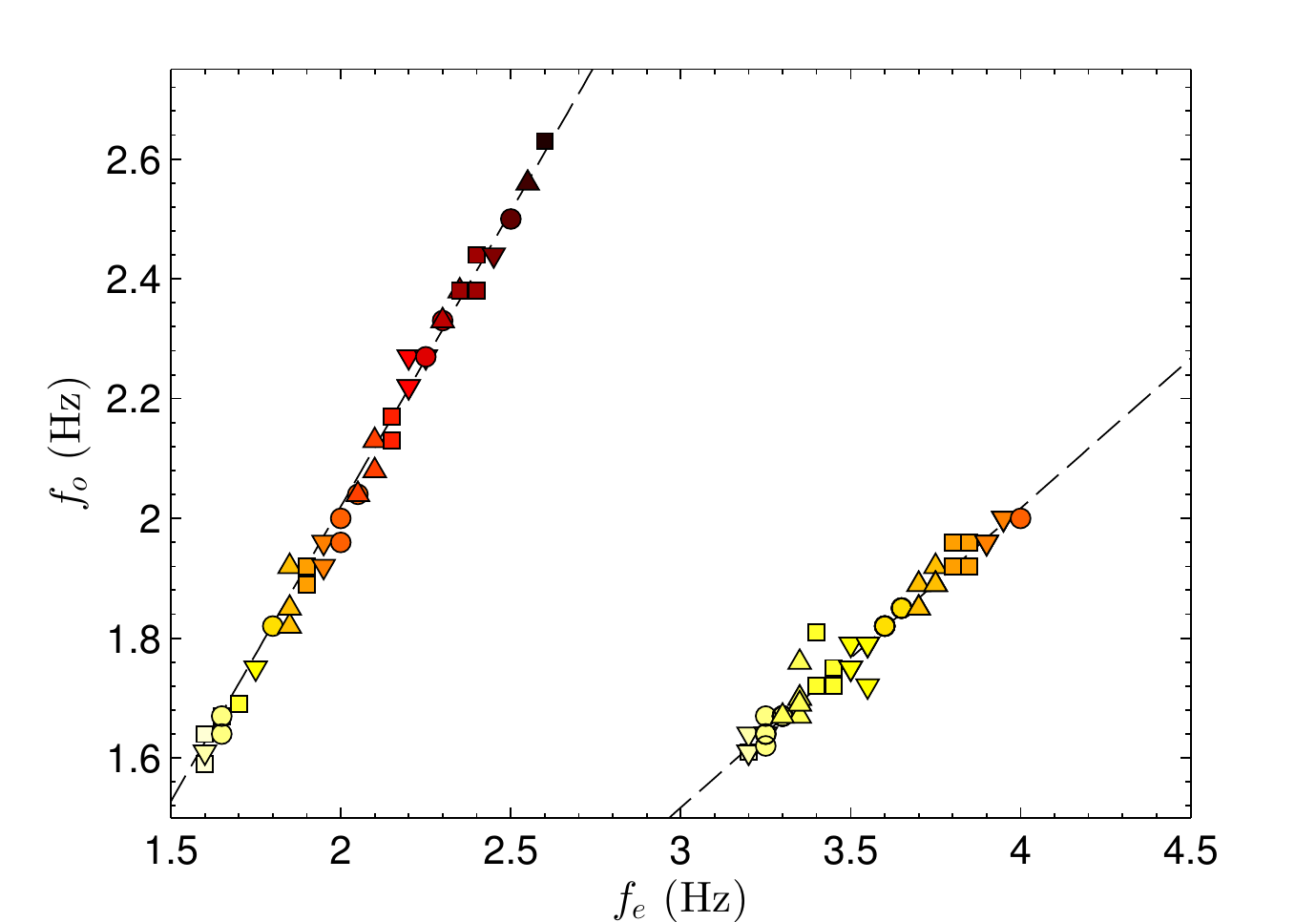}
	\caption{(Top) Phase diagram showing the appearance of vibration modes for different excitation parameters. The symbols correspond to the experimental data whereas the colored areas are the instability regions obtained numerically. (Bottom) Oscillation frequencies of the pendulum (denoted by $f_0$) versus the excitation frequency ($f_e$). On both graphics, each symbol corresponds to a vibration mode.}
	\label{fig:diagram}
	\end{center}
\end{figure}

 A phase diagram showing the domain of existence of vibration modes $n$ for different excitation parameters is presented in Fig.\ref{fig:diagram}. The numerical simulation of Eq. (\ref{motion}) evidences the existence of different sets of resonances. Due to the damping, there exist a threshold for the excitation of the different modes. The lowest threshold is obtained for the so-called 2:1 or subharmonic resonance, where the frequency of the excitation is twice the natural frequency of the pendulum. Whithin each resonance set, each mode has its own instability region, or Arnold tongues, which are represented in Fig. (\ref{fig:diagram}) with a different color for each mode. The dark shaded region corresponds to numerically unstable solutions. Note also the existence of a second set at lower frequencies, with a higher threshold, denoting the 1:1 resonance, that corresponds to modes excited when driving the system at the natural frequency of the pendulum $f_0=\sqrt{(g/L)}\approx 1.6$ Hz. 
 The symbols in Fig.(\ref{fig:diagram}) correspond to experimental data. From red to blue, the mode number increases, similarly to the numerical results.
No vibration modes are observed at frequencies below $f_e/f_0 = 0.95$, as there are no unstability regions predicted by the theory. From $f_e/f_0 =0.95$ to $1.90$, a first group of vibration modes
appears when the excitation amplitude is greater than a threshold. These modes oscillate at a
frequency equal to the excitation frequency ($f_\mathrm{osc} = f_e$), and correspond to the 1:1 parametric resonance of the system. From $f_e/f_0 =1.90$ to $f_e/f_0 = 3.17$, a second group of vibration modes oscillating at subharmonics of the excitation frequency
($2f_\mathrm{osc} = f_e$) is observed, corresponding to the 2:1 resonance. Therefore, two sets of resonance modes have been observed. Comparing the trend of simulation and experimental data shown in Fig.~\ref{fig:diagram}, we can conclude that there exists a good agreement.

It is important to note that pure modes exist only in a narrow region. Close to the parametric instability threshold, the Arnold tongues corresponding to each mode $n$ overlap. Thus, most of the modes observed actually correspond to a mixing of neighboring modes, with one mode being clearly dominant. Mode mixing is therefore expected to occur too in ultrasound-driven bubbles; however, such mode mixing has been postulated but not analyzed or observed in such microbubbles. The mechanical analog presented here could be used to explore these complex phenomena. 

Finally, we note that the maximum excitation has been measured at $f_e = 5$ Hz ($f/f_0=3.2$). Above this frequency, measurements cannot be performed due to the setup vulnerability. Oscillations becomes jerky and the coupling between pendula often breaks.

\section{Conclusions}
The interaction between a microbubble and an acoustical field has been studied
through the use of a macroscopic acousto-mechanical analogy. As for
real bubbles, vibration modes and modes mixing have been observed. It is known that, in the case of microbubbles, vibration modes display a strong subharmonic behavior. Here,
with the pendula ring, vibration modes have also been excited in the region where $f_\mathrm{osc} = f_e$, corresponding to a 1:1 resonance.

Localized modes like breather have not been discussed here but a previous study on the
pendula ring shows the appearance of such modes \cite{sanchez2013,sanchez2014}. One can thus expect to observe these oscillatory behaviors in microbubbles. These results allow to offer new insights in the
study of microbubble's dynamics. The analogy between the macroscopic behavior of the
pendula ring and the microscopic behavior of a microbubble presented here is valid for low
order modes (i.e. in the limit $na/R_0 \ll 1$) and provides important insights for the comprehension
of microbubble dynamics. Further studies are underway, particularly an expansion
of the analogy to consider encapsulated microbubbles to get information for both imaging
and therapeutic applications using contrast microbubbles.

\section*{acknowledgments}
Thanks to Jean-Marc Gr\'{e}goire and Jean-Yves Tartu for their fruitful help in the design of the experimental setup. Jennifer Chaline was the recipient of a PhD fellowship from the R\'{e}gion Centre (France). Financial support from the University Francois-Rabelais doctoral school of Tours was granted to finance a research stay on this topic at the Universidad Politecnica de Valencia (Spain). A. Mehrem is recipient of a research fellowship from Generalitat Valenciana (Santiago Grisolia program).  

\bibliographystyle{aipnum4-1}
\bibliography{bubble_bib}



\end{document}